# Intermediate magnetisation state and competing orders in $Dy_2Ti_2O_7$ and $Ho_2Ti_2O_7$

R. A. Borzi,[1,2] F. A. Gómez Albarracín,[2,3] H. D. Rosales,[2,3] G. L. Rossini,[2,3] A. Steppke,[4,5] D. Prabhakaran,[6] A. P. Mackenzie,[4,5] D. C. Cabra,[2,3] and S. A. Grigera[1,2,4,*]

[1] Instituto de Física de Líquidos y Sistemas Biológicos, UNLP-CONICET, La Plata 1900, Argentina, [2] Departamento de Física, Facultad de Ciencias Exactas,Universidad Nacional de La Plata, 1900 La Plata, Argentina, [3] Instituto de Física de La Plata, UNLP-CONICET, 1900 La Plata, Argentina, [4] School of Physics and Astronomy, University of St. Andrews, St Andrews KY16 9SS, United Kingdom  [5] Max Planck Institute for Chemical Physics of Solids, Nöthnitzer Str. 40, Dresden, Germany  [6] Department of Physics, Clarendon Laboratory, University of Oxford, Parks Road, Oxford OX1 3PU, United Kingdom,  * sag@iflysib.unlp.edu.ar

**Among the frustrated magnetic materials, spin-ice stands out as a particularly interesting system. Residual entropy, freezing and glassiness, Kasteleyn transitions and fractionalisation of excitations in three dimensions all stem from a simple classical Hamiltonian.  But is the usual spin-ice Hamiltonian a correct description of the experimental systems? Here we address this issue by measuring magnetic susceptibility in the two most studied spin-ice compounds, $Dy_2Ti_2O_7$ and $Ho_2Ti_2O_7$, using a vector magnet.  Using these results, and guided by a theoretical analysis of possible distortions to the pyrochlore lattice, we construct an effective Hamiltonian and explore it using Monte Carlo simulations.  We show how this Hamiltonian reproduces the experimental results, including the formation of a phase of intermediate polarisation, and gives important information about the possible ground-state of real spin-ice systems.  Our work suggests an unusual situation in which distortions might contribute to the preservation rather than relief of the effects of frustration.**

Spin-ice systems owe their name to an analogy of their ground-state construction rules to those in Pauling's model for proton disorder in water ice [1,2]. In their simplest variant, they can be described in terms of centre-pointing classical Ising spins at the binding points of a network of corner-shared tetrahedra forming a pyrochlore lattice (see Fig. 1a). In real materials, such as $Dy_2Ti_2O_7$ (DTO) and $Ho_2Ti_2O_7$ (HTO), this is an accurate description for the low temperature behaviour. It was recognised early that given the large magnetic moments of the rare-earth ions (of the order of 10 Bohr magnetons) in addition to a ferromagnetic nearest neighbour coupling $J$ (of the order of 1K) a dipolar term was necessary in the Hamiltonian in order to describe the experimental results [3-5]. The minimal model that became the norm in the description of spin-ice materials is the *standard dipolar spin-ice model* (s-DSM):

$$\mathcal{H} = J \sum_{<i,j>} \mathbf{S}_i \cdot \mathbf{S}_j + D r_1^3 \sum_{i,j} \left[ \frac{\mathbf{S}_i \cdot \mathbf{S}_j}{|\mathbf{r}_{ij}|^3} - \frac{3(\mathbf{S}_i \cdot \mathbf{r}_{ij})(\mathbf{S}_j \cdot \mathbf{r}_{ij})}{|\mathbf{r}_{ij}|^5} \right] \quad (1)$$

where $r_1$ is the nearest-neighbour distance, $\mathbf{r}_{ij}$ is the distance between spins *i* and *j*, $\mathbf{S}_i$ is a classical spin of unit length, and the dipolar constant, $D$, is of the same order than $J$. Spin-ice systems belong to the class of *geometrically* frustrated magnets. Frustration is possible in other ways, but the class of geometrically frustrated materials has the advantage of better experimental control and theoretical description than the case where disorder is needed. An inherent weakness of geometrical frustration is that it might be relieved by distortions to the lattice [6-8]. The spin-ice materials DTO and HTO are fairly robust to distortions; experimentally, the properties of the frustrated state and its low-lying excitations are seen to dominate the intermediate temperature regime [9]. However, as we shall see, the effects of small distortions in this system might account for some specific experimental features and determine the presence or absence of an ordered state at low temperatures.

The different limits of the classical s-DSM already contain the most singular and attractive features of spin-ice materials, notably residual entropy, emergent gauge structure, fractionalised monopolar excitations [10-12], and 2D and 3D Kasteleyn transitions [13-15]. The s-DSM also predicts an ordered state at very low temperatures [16]. As the field developed, it was soon recognised that additional interaction terms are needed to properly account for the experiments and an effort was made to determine an *empirical* Hamiltonian from the experimental data [17,18]. In particular, the most complete work of this type is that of Yavors'kii and collaborators [19]. In their approach, Monte Carlo simulations of a model with dipolar interactions, and isotropic first ($J_1$), second ($J_2$) and third ($J_3$) nearest neighbour exchange interactions are compared with DTO experimental susceptibility, specific heat and magnetisation data for zero magnetic field $H$ [20-22], with $H$ applied in the 112 direction [23,24], and with the empirical dependence of its polarisation transition with $H // 111$ [25]. By means of this comparison they put bounds to the values of $J_1$, $J_2$ and $J_3$ and arrive at an empirical Hamiltonian, the *generalised dipolar spin-ice model* (g-DSM). The g-DSM gives a very good phenomenological description of several of the spin-ice features, and reproduces the low temperature neutron scattering pattern of DTO noticeably better than the s-DSM [19].

Despite its successes, the g-DSM model does not fully reproduce the low temperature behaviour of the spin-ice materials. In this work we address two instances of these, which were unknown at the time the model was developed: the doubling of the polarisation transition with fields in the neighbourhood of 111 [26,27] and the recent report of an upturn in the specific heat below 0.6 K [28]. Based on our experimental results and those in the literature, we construct a new spin-ice model and show that in addition to reproducing the experimental features, it predicts an intermediate magnetisation state for fields near [111] and the existence of different types of possible order at low temperatures when no field is applied

## Results

**Magnetic Susceptibility.** We start by discussing the magnetic susceptibility at low temperatures when an external constant field is applied in the neighbourhood of the crystallographic [111] direction. For our experiments, single crystalline samples of DTO and HTO where used. The ac-susceptibility was measured in a dilution refrigerator with an external field applied using a triple-axis vector (see Methods).

The change in ac-susceptibility at the polarisation transition is plotted in Figure 1 for the HTO single crystal (analogous behaviour is seen in the DTO sample) as a function of field at a fixed temperature of $T$= 0.18 K for different angles from the [111] direction (see Fig. 1b). Fig. 1c shows field rotations from [111] towards [110], while Fig. 1d shows rotations towards [112]. The usual spin-ice models predict a single transition between the partially frustrated Kagome-ice state at low fields and the polarised state at high fields. In the experiments this is true when the field is perfectly aligned with [111], or when it is rotated towards [110], but the transition splits into two peaks, with an intermediate polarisation region, when the field is rotated towards [112]. This behaviour is seen more clearly in Figure 2, where the change in susceptibility is shown for both materials as an interpolated contour plot. In this figure it is also quite noticeable that the angular dependence of the lower transition field is quite small when the field is rotated towards [112]. Similar behaviour had been previously observed in magnetisation experiments in DTO [26].

The failure of the usual spin-ice Hamiltonians to reproduce an experimental feature common to both spin-ice materials suggests that there is an additional physical mechanism that is being neglected in the theoretical treatment. In the following, we argue that a missing ingredient can be the effect of distortions in the magnetic couplings of the pyrochlore lattice

**The role of distortions.** We analyse how the s-DSM Hamiltonian (eq. 1) is modified in the presence of distortions. We consider the simplest case: classical Einstein phonon modes in the pyrochlore lattice, parameterized as $\mathbf{u}_i \approx \mathbf{r}_i - \mathbf{r}_i^0$, with a linear dependence of the exchange couplings on distances: $J(\mathbf{r}_{ij}) \approx J_0[1 - \alpha \, \mathbf{r}_{ij}^0 \cdot (\mathbf{u}_j - \mathbf{u}_i)]$, where $J_0$ is the undistorted exchange constant and $\alpha$ is the spin-phonon coupling constant, and a simple quadratic term for the energy cost of distortions. Expanding the variation of $\mathcal{H}$ up to linear order in $\mathbf{u}_i$ one obtains a *magnetoelastic* Hamiltonian that can be written compactly as:

$$\mathcal{H}_{\mathrm{me}} = \mathcal{H} + \sum_i \frac{K}{2}(\mathbf{u}_i)^2 + \sum_i \mathbf{u}_i \cdot \sum_{j=N(i)} \mathbf{F}_{ij} \quad (2)$$

where $K$ is the elastic constant for classical Einstein phonons, $N(i)$ stands for the neighbours of site $i$, and $\mathbf{F}_{ij}$ encodes the quadratic spin interactions between a site $i$ and a neighbour $j$.

An effective Hamiltonian can be obtained from $\mathcal{H}_{\mathrm{me}}$ by integrating out the phononic degrees of freedom [29] (see Methods). This effective Hamiltonian can be cast into a simple form where the effect of distortions to the lattice is translated into a change in the nearest neighbour interaction and the emergence of further neighbour interactions. These interactions have their origin in the local changes in the magnetic interaction due to changes in the relative positions of the lattice sites, and are thus strongly dependent on the lattice connectivity between neighbours. Further neighbours of equal distance but connected to the original site by a different number of lattice sites develop different interaction terms. In the case of the pyrochlore lattice (see Fig. 1a) this mechanism provides two kinds of third-nearest-neighbour interactions, $J_3$ and $J_3'$.

**The model.** While it cannot be expected that this simple analysis based on classical phonons will give accurate values for the effective exchange constants of the real system, which consists of a far more complicated lattice once all the non-magnetic atoms are taken into account, it provides the key observation that, regardless the specific material, further neighbour interactions should be considered and distinction made between differently connected neighbours of the same order, even if these terms or differences were negligible from the exchange integrals. With this information, we constructed a model to third nearest neighbours that incorporates these distinctions, the d-DSM, and explored it using Monte Carlo simulations (see Methods and Supplementary Note 1). As additional constraints, we keep $J_2$ and the mean value $(J_3+J_3')/2$, within the first set of intervals determined by Yavorsk'ii et al. for the g-DSM [19] in order to retain compatibility with the experimental results this model was fitted to (see also Supplementary Note 2 and Supplementary Figure 1).

Figure 3 shows the magnetic susceptibility for the d-DSM, with $J_1$=3.41K, $J_2$=0, $J_3$=-0.02K, $J_3'$=0.07K and $D$=1.3224K (Fig. 3a), and for the g-DSM model (Fig. 3b), both as interpolated contour plots. Unlike the s-DSM or g-DSM, and in keeping with the experiments, the susceptibility calculated in the d-DSM has a flatter angular dependence of the critical field for rotations towards [112], showing a broad peak that eventually resolves into two peaks, and remains unique when the field is rotated towards [110], all of which is in correspondence with the behaviour shown in Figure 2 for the experimental results in the neighbourhood of [111]. Figure 3b shows the best possible fit to the data using the g-DSM: by forcing the parameters towards the higher end of the intervals given in [19], it is possible to eventually achieve a small doubling of the transition when rotating towards [112], but it completely fails to reproduce the angular dependence of the critical field. The addition in the d-DSM of *different types* of $J_3$ is therefore necessary to be able to reproduce the susceptibility experiments close to [111], even at a qualitative level. Further terms and analysis are required to reproduce in detail the full angular dependence.

**Intermediate phase.** A known consequence of the coupling between spin degrees of freedom and distortions is the stabilization of plateaux in the magnetisation (see e.g. [30] and [31]). It therefore comes as little surprise that the doubling of the polarisation transition corresponds, in the limit of low temperature and perfectly homogeneous tensions, to the presence of a plateau between the kagome-ice and the fully polarised state (see Figure 4a). This plateau marks an intermediate ordered state where the ice rule is broken in only half of the tetrahedra. In order to visualise these states, it is helpful to mark by a red or blue dot the defect or excess of inward pointing spins. Defects of different colour interact approximately as charges of opposite sign [11]. Full polarisation corresponds in this picture to a complete checkerboard pattern of blue and red dots. The intermediate states of Figure 4 are easily pictured as alternate arrangements of stripes of the checkerboard pattern separated by empty stripes (see Figure 4).

**The ground state.** One of the remarkable predictions of all dipolar spin models is that, contrary to initial expectations, spin-ice materials may have no extensive residual entropy even in the absence of an applied magnetic field. As shown by Melko *et al.* [16], the introduction of dipolar interactions leads to a sharp transition at approx. 0.18K into an ordered ground state. At present there is no direct experimental determination of the ground-state of spin-ice systems, since below 0.6 K the characteristic relaxation time of the system seems to grow faster than exponentially [32,33] and most measurements probe properties of the system out of equilibrium [34]. A recent experiment [28], in which the zero field low temperature specific heat was measured over much longer timescales, reports the possible observation of a peak in the specific heat that might correspond to the onset of order in DTO. However, the temperature at which this onset is seen (approx. 0.6 K) is not compatible with the prediction of Melko et al. for the s-DSM. The situation is very similar for the g-DSM: the additional interaction terms of this model leave the ordered state unaltered, and the specific heat (by construction), shows no upturn down to 0.2 K.

The model presented in this work (the d-DSM) is a different scenario altogether. In the simple homogeneous case, and within the constraints on the parameters imposed by the experiments (crucially our susceptibility near [111]) two different ordered ground states are possible depending on the relation between the exchange constants.

The Figures 5a and 5b show cuts along [100] of the two ordered states obtained at low temperatures for different values of the exchange constants. For small values of $J_2$, the relevant parameter that switches between the two types of order is the difference, $\Delta J_3 = J_3' - J_3$. The scale in the middle (Fig. 5c) indicates, within the region compatible with the experiments (shaded in blue), the values of $\Delta J_3$ (for fixed average $J_3$, and $J_2 = 0$) where each type of order is seen. As expected, for low values of $\Delta J_3$, the ordered state is identical to the one originally found in the s-DSM. A simple description of this state (state I) is to consider it an antiferromagnetic arrangement of ferromagnetic chains of spins (indicated with the same colour in the figure) at axes rotated $\pi/4$ with respect to the original lattice. The second state (state II) can be thought as a simple extension of state I where the unit blocks are now *pairs* of ferromagnetic spin chains (also indicated with the same colour in the right hand side figure) ordered antiferromagnetically. In both cases, the non-shaded region in the figure corresponds to the unit cell.

These are the two states of lower energy that have been identified depending on the degree of asymmetry in the two $J_3$. It is straightforward to conjecture that there is a wealth of possible intermediate states with energies differing by a few mK (in $1/k$), constructed from all the possible periodic intercalations of different numbers of alternating single and double chains in each direction, or even of the alternation of bundles of a higher number of chains. This by itself has already the consequence that it will be extremely difficult for any experiment to probe equilibrium properties, but the situation in a real sample is even more complicated: the most likely experimental state is that the sample will have a distribution of tensions and distortions within its volume. This will lead to local changes in $\varDelta J_3$, coming from the distortion-induced contribution to the two $J_3$, which in turn will result into local energetic rearrangements of the possible states. No long-range ordered state will be favourable in the whole sample. The effect of tensions and distortions could also be intimately related with the experimental observation of freezing or glassy behaviour below temperatures of about 0.6 K [35-38].

**Specific heat.** Figure 5d shows the zero-field specific heat divided by temperature, calculated for the d-DSM with the same values used for Figure 3, compared with the experimental data for DTO from [12] and [28] (a finite size analysis of the simulated specific heat is shown in the Supplementary Note 3 and Supplementary Figure 2). Both experimental sets of data coincide at high temperatures, down to about 0.6 K below which the characteristic time of the system becomes extremely long and the results from [28], taken over longer times, are believed to be closer to equilibrium. The upturn seen in this data has been interpreted as a possible sign of onset of long-range order. The specific heat calculated from the d-DSM matches the high temperature behaviour of the experiments and shows a sharp peak at a temperature close to 0.2 K, which corresponds to the ordering transition into state I or II (depending on the value of $\varDelta J_3$). It would be tempting to associate this peak with the upturn of [28], and match the two sets of data by making small adjustments to the exchange constants of the d-DSM. However, as discussed before, it might not be realistic to expect a homogeneous long-range ordered state in a real sample, but rather a wide range of local configurations and a multitude of energetically close excitations. In this case, features matching those seen in the specific heat data of [28] have their origin in a Schottky-type peak consequence of this proliferation of states of very similar energies in this range of temperatures. A more definitive answer to this question would come from neutron scattering experiments performed under the same relaxation conditions as the specific-heat experiment.

**Discussion**

In summary, in this work we present experiments performed in the two paradigmatic examples of spin-ice materials, DTO and HTO, where we have measured the angular dependence of susceptibility using precise alignment with the crystallographic axes, concentrating in particular in the polarisation transition with field in the neighbourhood of [111]. We show that a generic feature of these two materials, absent from any theoretical spin-ice model in the literature, is a marked anisotropy in $\chi$, which shows a doubled transition when the field is rotated towards [112]. This doubled transition had been previously observed in specific heat and magnetisation experiments in DTO [23,26], but not in HTO. We considered the possibility of distortions, a missing ingredient from the usual spin-ice models, and by performing the simplest possible analysis on their effect on the standard dipolar spin-ice model we found that they provide a justification for the addition of interactions between second and two kinds of third nearest neighbours. This is a generic feature of distorted spin-ice materials, and applies even if the exchange terms for the specific material were negligible. There is a crucial difference in the current treatment of thermal vibrations to what already exists in the literature, which is a consequence of the presence of long-range interactions in this class of materials. Thermal vibrations usually just renormalize constants, but in the present case dipolar long-range interactions lead to the appearance of couplings that would not exist in the absence of vibrations. Thus, our treatment of the simple d-DSM leads to a more complex Hamiltonian, that resembles that of e.g. ref. [19], but is not obtained in a purely empirical manner.

Provided with these ingredients we constructed a new model (the d-DSM) and explored its characteristics using Monte Carlo simulations. The new model reproduces the anisotropy found in the experiments with fields in the neighbourhood of [111], and, despite the fact that it was not built based on neutron scattering data or *Cv*, the *S(q)* calculated for the d-DSM reproduces the experimental results (see Supplementary Note 4 and Supplementary Figure 3) and we have seen that the same is true for *Cv* (Fig. 5). More importantly, it predicts an intermediate magnetisation state, thus explaining the hitherto unresolved issue of a doubled polarisation transition with field close to the [111] direction, and also predicts the existence of different types of possible order at low temperatures when no field is applied. These results from the d-DSM are valid regardless the ultimate origin of the different interaction terms. The possibility that these further order terms could be a consequence of distortion would mean that the small tensions present in real samples could result in the absence of long-range order at low temperatures. This is probably the most surprising conclusion of this work, that distortions, which usually result in a relief of frustration, might be the reason why the system would remain disordered at very low temperatures.

Very recently, Henelius et al. [39] have presented an empirical spin-ice Hamiltonian constructed to describe three main sets of experiments in $Dy_2Ti_2O_7$ (the structure factor *S(q)*, the specific heat *Cv* and the magnetisation for *H* near [112]). The model Hamiltonians presented in both studies have several points of contact. In particular, both models predict the same possible ordered ground states for *H*=0 (those described in Fig. 5) and, crucially, the likelihood of a disordered ground state as a consequence of internal stress (in our case given by external strain, in theirs by impurities) and quasi-degeneracy. The similarity of these findings is remarkable taking into account that the core experiments used in their work and in ours are very different.

## Methods

**Experiments**

For our experiments, single crystalline samples of DTO and HTO where grown using an optical floating zone in St Andrews and Oxford, and they where oriented and cut into prisms of approximate dimensions (0.7 x 0.7 x 3) mm$^3$, with the long axis oriented along [111]. The ac-susceptibility was measured in a dilution refrigerator using two sets of detection coils, positioned so as to avoid cross-talk effects, and a drive coil. Each detection coil set comprises of two coils, connected antiparallel to each other. The drive coil is concentrically wound around the pick-up coils. The samples were placed in each of the upper pick-up coils and thermally grounded to the mixing chamber through silver and copper wires. A low-frequency and very low-amplitude excitation field of about 3 x 10$^{-5}$ T rms was generated by the ac current in the drive coil, and the response detected by a lock-in amplifier. For accurate control of the external magnetic field, we used a bespoke triple-axis vector magnet capable of producing 9/1/1 T along z, x , and y. Careful centring using the vector field capability allowed $\theta$ to be determined to an accuracy of 0.1 degree relative to the crystallographic axis [111].

**Theoretical analysis of the effect of distortions**

A simpler effective Hamiltonian can be obtained from $\mathcal{H}_{me}$ (eq. 2.) by integrating out the phononic degrees of freedom [26]. In the spin-ice model $\mathbf{S}_i = \sigma_i \mathbf{e}_i$, with $\sigma = \pm 1$, which leads to important simplifications: since $\sigma_i^2 = 1$, the usual quartic terms become quadratic. Still, due to the long-range nature of dipolar couplings one encounters terms $\mathbf{F}_{ij}$ of any range, and the analysis of the effective magnetic model is rather complex. In order to compare the predictions of our model with those in the literature we have considered an expansion of the corrections to the full effective magnetic Hamiltonian up to third neighbours.

The effective magnetic model obtained reads

$$H_{\text{eff}}/J = J_1^{\text{eff}} \sum_{<i,j>_1} \sigma_i \sigma_j + J_2^{\text{eff}} \sum_{<i,j>_2} \sigma_i \sigma_j + J_3^{\text{eff}} \sum_{<i,j>_3} \sigma_i \sigma_j$$
$$+ J_3'^{\text{eff}} \sum_{<i,j>_3} \sigma_i \sigma_j + \text{further dipolar terms}$$

where $<i,j>_r$ indicates $r$-th range neighbours. The couplings up to third neighbours are

$$J_1^{\text{eff}} = -\frac{1}{3} + \frac{5}{3}\frac{D}{J} - \delta, \; J_2^{\text{eff}} = -\frac{D}{3\sqrt{3}J} + \delta, \; J_3^{\text{eff}} = -\frac{D}{8J} + 2\delta, \; J_3'^{\text{eff}} = \frac{D}{8J}$$

where $\delta$ is the contribution arising from the phonons, $\delta = J/4Ka^2(\alpha/3J - 5D/J)^2$. Further range terms are the unmodified dipolar interactions.

The first observation is that the crystal environment distinguishes two different third neighbour couplings in $H_{\text{eff}}$: $J_3^{\text{eff}}$ and $J_3'^{\text{eff}}$ distinguish between third nearest-neighbours connected through the lattice by a minimum of one or two other lattice sites respectively. One of them is sensitive to distortions, while the other is not. A second observation is that the relative strengths of all couplings in $H_{\text{eff}}$ depend only on two microscopic parameters: $D/J$ and $\alpha/J$. Corrections of longer-range magnetic couplings can be done systematically, depending on the same microscopic parameters. We have checked that fourth order corrections do not change the main results of this work.

It is not realistic to expect that a simple classical model such as the one presented here will give an accurate description of the real system at low temperatures. The presence of soft modes (speculated for spin-ice materials in [40] or seen in Tb$_2$Ti$_2$O$_7$ [41]) could be underlying the relative success of such a treatment, and is worthy of being investigated.

**Monte Carlo simulations**

For the d-DSM model we used the following Hamiltonian

$$\mathcal{H}_d = J_1 \sum_{<i,j>_1} \sigma_i\sigma_j + J_2 \sum_{<i,j>_2} \sigma_i\sigma_j + J_3 \sum_{<i,j>_3} \sigma_i\sigma_j + J_3' \sum_{<i,j>_{3'}} \sigma_i\sigma_j$$

$$+ Dr_1^3 \sum_{i,j} \left[ \frac{\mathbf{S}_i \cdot \mathbf{S}_j}{|\mathbf{r}_{ij}|^3} - \frac{3(\mathbf{S}_i \cdot \mathbf{r}_{ij})(\mathbf{S}_j \cdot \mathbf{r}_{ij})}{|\mathbf{r}_{ij}|^5} \right] - \mu \sum_i \mathbf{S_i} \cdot \mathbf{B}$$

where the <i,j>$_3$ and <i,j>$_{3'}$ distinguish between third nearest-neighbours connected through the lattice by a minimum of one or two other lattice sites respectively (see Figure 1a).

We performed Monte Carlo simulations, using Ewald summations to take into account long-range dipolar interactions [42,43] in a conventional cubic cell for the pyrochlore lattice. We simulated systems with $L\times L\times L$ cells with periodic boundary conditions. Basic parameters were taken from reference [19], including cell size, lattice parameter and magnitude of magnetic moments. Thermodynamic data as a function of field were collected using a variation of single spin-flip

Metropolis algorithm. For Figure 3 we chose a temperature of 0.4 K to avoid problems associated with hysteresis. For the specific heat data below 0.6 K we mixed the usual Metropolis with a dynamics that conserves defects for a certain number of Monte Carlo steps before letting the system relax (the algorithm is a variation of the one used in [43] and is described in the Supplementary Note 1). In the case of magnetisation measurements at 0.1 K (Figure 4) data at each field was collected after a slow annealing, averaging over five independent runs.

The average value of the third neighbour interactions was kept within the range specified in reference [19] to retain compatibility with previous experiments in DTO. As an example, with the values of $J$ used in figures 3-5, we studied the transition to ferromagnetic ordering with field nearly along the [112] direction [19, 21]. We obtained a transition temperature of ≈0.35K which could be justified considering angular deviations of less than 0.5 degrees off perfect alignment.

Regarding the double peak near [111], we contrasted our simulations with the static magnetisation data obtained in [26], which shows narrower features than the ac-susceptibility. The range fixed in Figure 5 was determined looking for special key details: i) a magnetisation increase of only ~$0.025\,\mu$ when the field is tilted from 0 to -5 degrees (towards [112]) at a fixed field of 0.9 T, ii) an increment for the first critical field of less than 0.06 T when tilting the field from 0 to -5 degrees; and iii) a difference between the two critical fields at -5 degrees of ≈0.1 T.

**Data availability.** All relevant data are available from the authors.

**Acknowledgements**

We thank R. Moessner, C. Castelnovo and M. Gingras for helpful discussions, and the financial support of ANPCYT through PICT 2013-2004 and PICT 2014-2618 and CONICET (Argentina), the EPSRC and the Royal Society (UK).

**Author contributions**

D.P. prepared the HTO crystal samples. R.A.B, S.A.G. and A.S. designed and performed the susceptibility measurements. F. A., H.D.R., D.C. and G.R. did the theoretical analysis, R.A.B. performed the Monte Carlo simulations The manuscript was initially drafted by S.A.G., and optimized by S.A.G, R.A.B, D.C. and A.P.M with input and discussion from all authors.

**Additional Information**

**Supplementary information** accompanies this paper at http://www.nature.com/naturecommunications

**Competing financial interests:** The authors declare no competing financial interests.

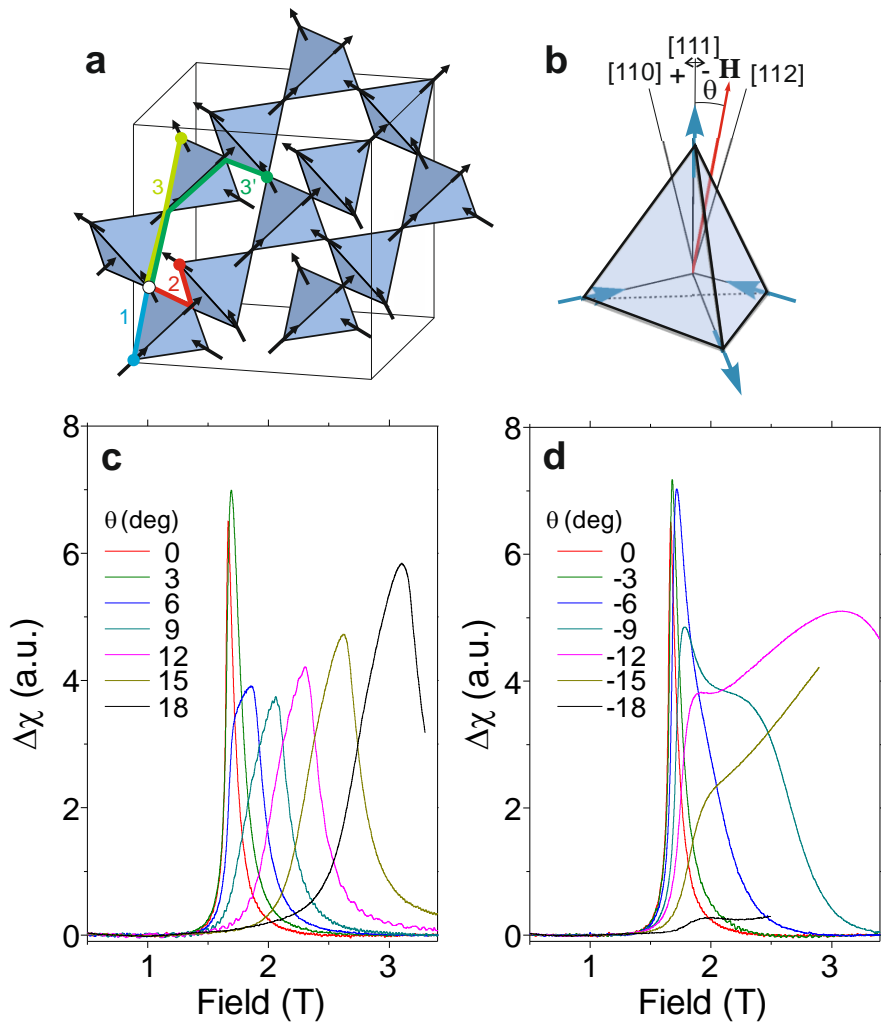

**Figure 1 | Magnetic susceptibility at the polarisation transition.** The data shown are from $Ho_2Ti_2O_7$, but are representative of both materials. (**a**) A schematic view of the pyrochlore lattice: spins are represented as black arrows, and tetrahedra are coloured light blue. The coloured lines show, starting from the white circle, a first (light blue), second (red) and two types of third (green and light green) nearest neighbour. (**b**) Sign convention for the angle shown in a single tetrahedron: rotations towards [112] are negative while those towards [110] are positive. (**c**) Magnetic susceptibility: when the field is rotated towards [110] the peak remains unique. (**d**) As the field is rotated towards [112], the peak splits into two, the upper one moves faster with field as the angle is increased. The temperature for all curves is fixed at 0.18 K and in all cases a smoothly varying background has been subtracted for clarity

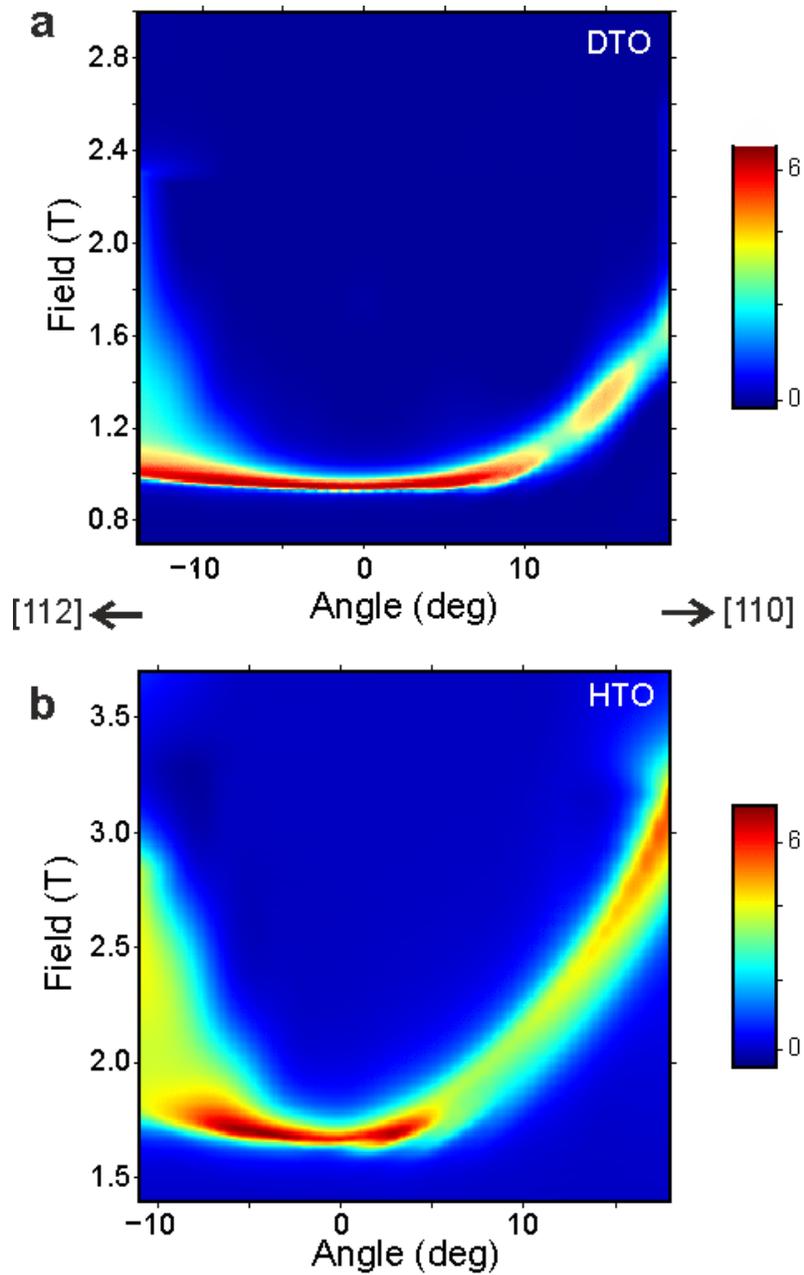

**Figure 2| Contour plots of the magnetic susceptibility for DTO and HTO.** (a,b) Interpolated contour plot of the magnetic susceptibility, for DTO and HTO respectively, at $T$ = 0.18 K. Each graph is constructed based on approximately 40 traces of $\chi$ versus angle at fixed temperature and field. Positive (negative) angles correspond to a rotation towards [110] ([112]). For both materials the transition widens –eventually resolving into two peaks- as the field is rotated towards [112]. The effect is more pronounced for HTO. The origin of this difference could lie on the non-Kramers nature of $Ho^{3+}$ which makes it more susceptible to its local environment.

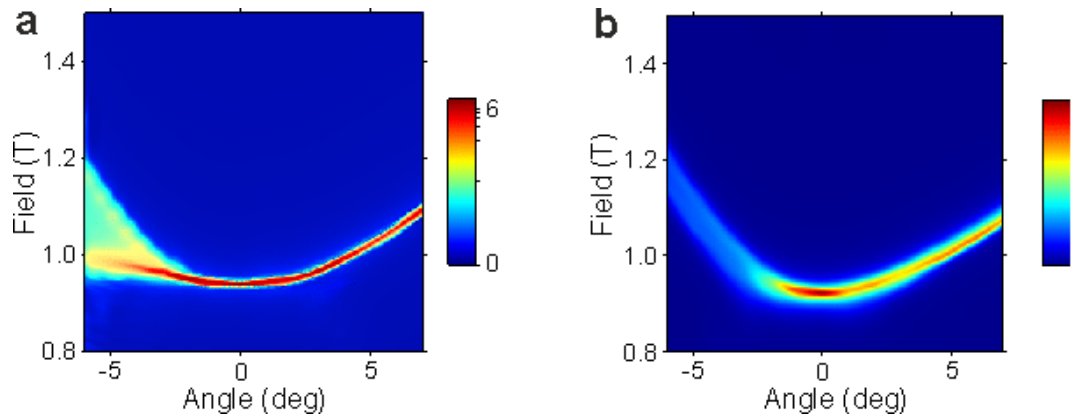

**Figure 3| Magnetic susceptibility from MC simulations.** (**a**) Contour plot of the susceptibility as extracted from the d-DSM, which shows a doubling of the transition and a marked anisotropy similar to that seen in the experiments (compare with previous data and ref. [26]). (**b**) The best possible fit to the data achievable with the g-DSM. In both cases the colour scale is non-linear in order to avoid saturation from the high intensity peaks.

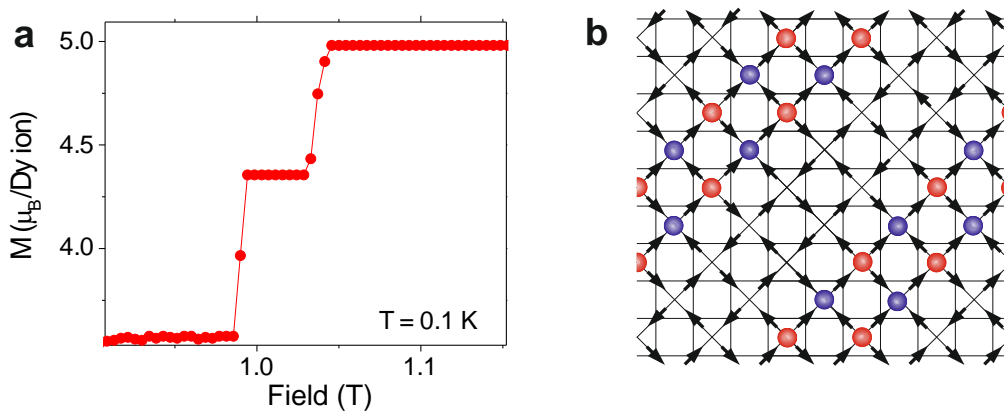

**Figure 4| Plateu in the magnetisation and intermediate state of the model with distortions (d-DSM).** (**a**) The simulated magnetisation curve in a homogeneous system as a function of field for T = 0.1 K with the field at an angle of -5 degrees from [111]. A mid-plateau is stabilised between the kagome-ice and the fully polarised state. (**b**) A sketch of this intermediate state projected along [100]. The coloured circles mark tetrahedra where the ice-rule is broken either by excess (blue) or defect (red) of inward pointing spins.

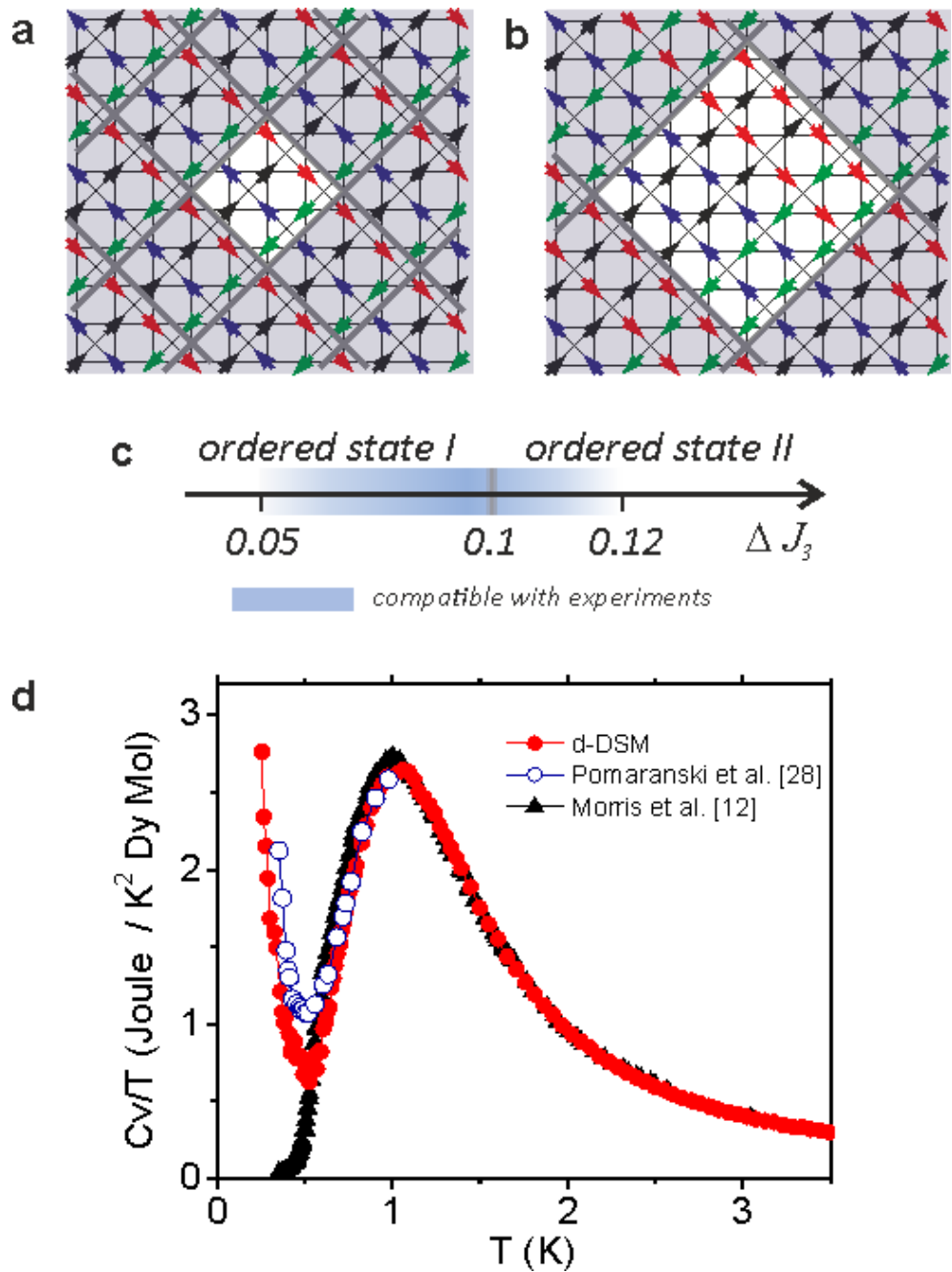

**Figure 5| Ordered states of the model with distortions (d-DSM).** Within the range of parameters fixed by the experimental susceptibility data, and depending on the difference $\Delta J_3$ between the third nearest neighbours exchange constant, the system orders into two possible states. **(a,b)** Cut along [100] of the two ordered states. The non-shaded regions correspond to the unit cells. **(c)** Ranges in $\Delta J_3$ where each state is observed. **(d)** The signature of this transition in the specific heat divided by temperature, compared with the results obtained for DTO in refs [12] and [28].

# Supplementary Figures

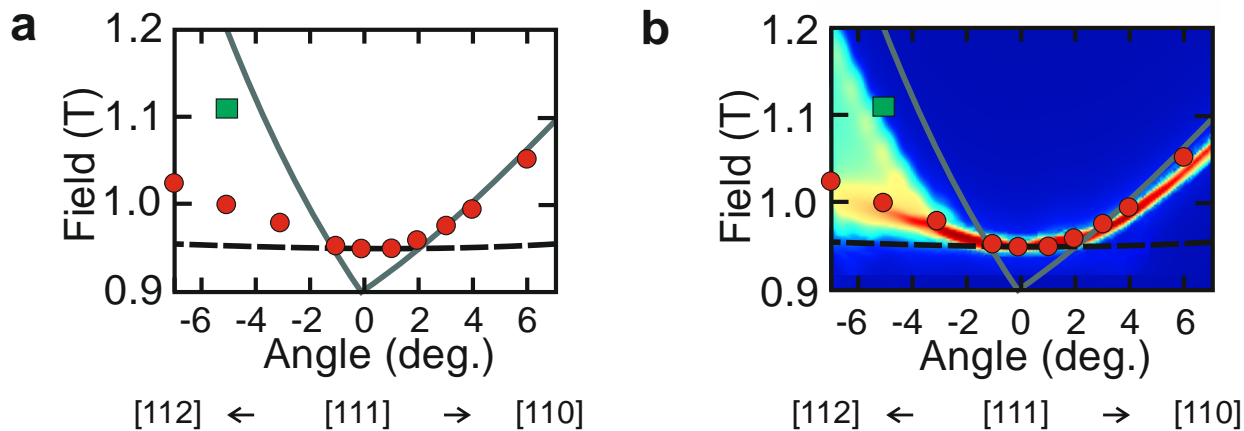

**Supplementary Figure 1 | Comparison between the experimental field-angle phase diagram and the g-DSM model.** **(a)** The phase diagram from Sato et al. [4] determined using magnetisation experiments (data from inset of Fig. 4 of [4]). The symbols represent the critical field measured with circles (red) for $Hc_1$ and the square (green) for $Hc_2$. The solid line corresponds to a hypothetical transition line, as calculated from the nearest neighbours model. **(b)** The diagram in **(a)** has been overlaid on top of a contour plot of the susceptibility calculated from the d-DSM (data from Fig. 3 of the main text). We can see that there is good coincidence between the experiments and the model.

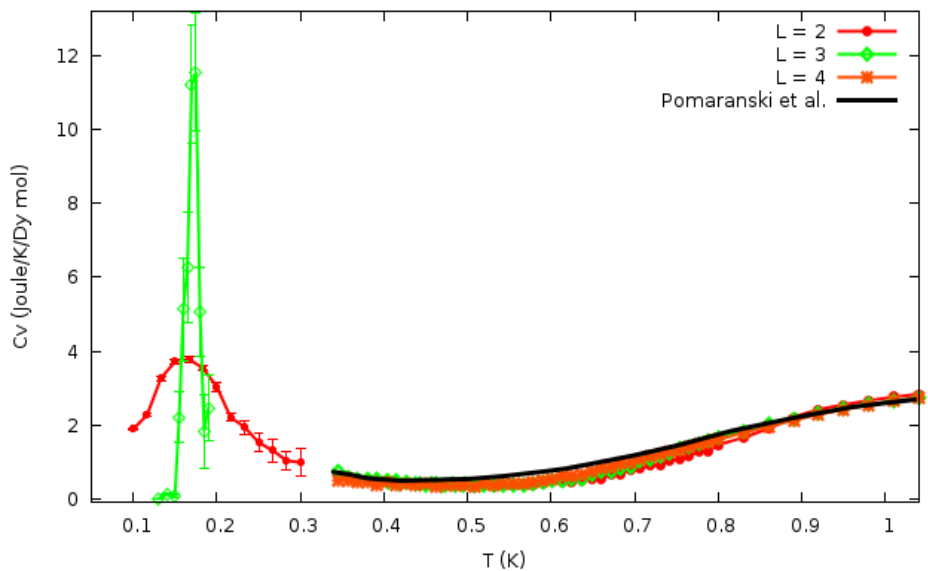

**Supplementary Figure 2 | Finite size scaling of the specific heat in the g-DSM model.** Specific heat calculated for the g-DSM model for two temperature ranges and different system sizes as indicated in the figure. The black line corresponds to the experimental data by Pomaranski et al. [5]

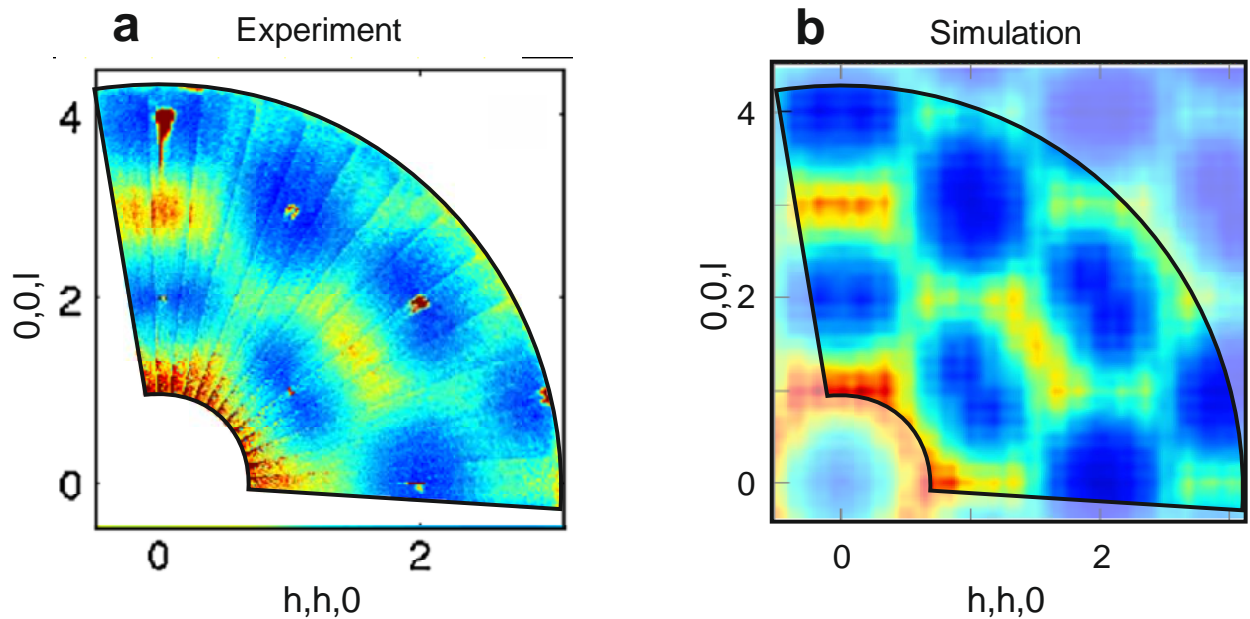

**Supplementary Figure 3 | Comparison between the experimental and simulated structure factors.** (**a**) the experimental $S(q)$ of $Dy_2Ti_2O_7$ as a function of the wavevector (from [7]) and (**b**) the $S(q)$ calculated from the d-DSM, both at 300mK.

# Supplementary Notes

### Supplementary Note 1: Monte Carlo simulations

In order to simulate the specific heat at low temperatures we have implemented a single spin flip dynamics, but different from the conventional in that we use of the Conserved Monopoles Algorithm (CMA) to reach equilibration at low temperatures in reasonable simulation times. An advantage of the CMA is that it can also be used in the presence of a magnetic field. A description of this algorithm can be found in [1] where it was introduced, and subsequently it has also been discussed and applied by other authors (see e.g. [2]). The CMA works in a statistical ensemble of conserved number of defects to the ice rule, which are free to propagate in the system. It can be shown that even a negligible density of defects (two in a lattice of thousands of sites) is enough, by lowering the barriers between states, to speed up the simulations substantially. For example (as shown in [1]) the CMA applied to the DSIM at zero field reproduces the results of Melko et al. [3].

We introduced in this work a small improvement to the CMA algorithm. In order to avoid any interference of the artificial defects in the values of the thermodynamic variables, before collecting one sample we remove the constraint and allow the defects to disappear while we let the system evolve for some steps. After the data is collected, we reintroduce a pair of defects and let the system continue its evolution. The model is hybrid, in that it only introduces the CMA after checking that the density of monopoles at a given temperature falls below a threshold. We have checked the code by reproducing again the phase transition towards the order state observed at 180 mK in the dipolar model by Melko and collaborators [3].

## Supplementary Note 2: Selection of the exchange constants for the model.

It is quite remarkable that a simple model for distortion gives an expression that, for a given $\delta$ and $J_1$, provides values for $J_2$ and for both $J_3$'s that lead to a Hamiltonian compatible with most of the experiments discussed in ref. 17 where the $J$ values were chosen in an ad-hoc fashion. This can be seen noting that the values of $J_2$ and $J_3^{av} = (J_3^{1} + J_3^{2)})/2$ predicted by the theoretical analysis discussed in the main text can be made to fit within the first range delimited in ref. 17: -0.20 K < $J_2'$ < 0 K; 0.019 K < $J_3'$ < 0.026 K. (Note that due to a geometrical factor implicitly taken into account in our model, our nominal value of $J_2$ has the opposite sign and with a modulus three times bigger than that in ref. 17, i.e. $J_2' = -1/3\, J_2$). By choosing a bigger $\delta$, we could even find a modest peak splitting for negative $\theta$ and no splitting in the opposite direction, for a limited value of $\theta$.

As a further refinement, and recognising that ours is just a minimal model, we used the parameters quoted in the text, which preserve the main constraints: nonnegative values for $J_2$ and $J_3^{av}$, with different $J_3^{1}$ and $J_3^{2}$, but which depart from the calculated relations. These parameters increased the peak splitting, extended the range of theta in which the double peak was observed (up to $\theta \sim 8$ deg in the simulations), and improved the quantitative comparison of our magnetisation curves with those in [4] (the low value of the magnetisation at fields just below the jumps was a key piece of information). We included, for comparison, a figure (supplementary Figure 1) that shows the experimental phase diagram determined measuring magnetisation by Sato el al. [4] and the same diagram overlaid on top of the susceptibility contour plot of the d-DSM (data from our Fig. 3 in the main text). Note that our sign convention for the angle is opposite to that of Sato et al. [4].

One (apparent) drawback we observed was that while we improved the similarity of the simulated double peak feature with the experimental data, the new set of *J*'s led to an unexpected increase of *Cv* at low *T*. Afterwards we realised that the new feature near ~0.4K in *Cv* was in agreement with that experimentally observed in ref. 1. It may be worth repeating that no effort was made to tune the *J* values in order to make our specific heat look like that in ref. 1 (see e.g. Figure 1 in section 3 below). Furthermore, as can be seen in section 4 below, this set of exchange constants also gives a very good account of the experimental *S(q)*.

## Supplementary Note 3: Finite size scaling of the specific heat at zero field.

We provide a finite size effect (FSE) study at *H=0*. We constrained it to small sizes, due to the computer time demanded by this study (see Supplementary Figure 2)

We concentrate on two important temperature ranges:

   a) Temperatures above ~ 0.4 K: This range is the main focus of our work. Recent experimental data for the specific heat is available down to a temperature of approximately 0.36K [5] (the lowest temperature of the data in Fig. 5 of the manuscript). Since we compare it with our equilibrated data, it is an implicit assumption of our work that this experimental specific heat is properly equilibrated.

It is reasonable to wonder if the rise we observe in $Cv/T$ in our simulations (bottom of Fig. 5 in the paper) is due to the imminent phase transition and thus affected by finite size effects. If so, the coincidence between our simulations and the experimental $Cv/T$ would be merely accidental. The figure above, including data for $L = 2, 3, 4$ and the experimental data from [5] shows that in the regime where $Cv$ turns up (between ~ 0.5 and 0.36 K) the variations due to finite size effects are not very significant and it is sensible to compare our simulated data with the experimental results. (A small finite size effect between 0.5 and 0.8 K may indicate that the coincidence can be even better in the thermodynamic limit).

b) Though we have been interested in the possible ground states predicted by our model, the phase transition occurring at low temperature has yet no experimental counterpart, and, as we mentioned, its nature is not one of the main concerns of this work. The figure shows some preliminary low temperature data, which are compatible with a first order transition. Indeed, the maximum of $Cv$ near 0.17 K increases, within error, like the volume of the system when going from $L=2$ to that $L=3$.

Measuring the specific heat with Ewald summations to take into account dipolar interactions, and using the conserved monopoles algorithm at low temperatures for $L =3$ with the statistics observed in the figure below (green curve) demanded 35 copies of a program running for a month (35x30x24 h = 25200 hours of computer time). To be more specific, it demanded 11 temperature points with $3\times 10^7$ MCS per point, per independent run. Given time and computer time limitations, we have not tried to measure the finite size effect for $L =4$ or bigger lattices at very low temperature.

### Supplementary Note 4: Calculation of the structure factor $S(q)$

We calculated the structure factor predicted by the d-DSM model introduced in this work. We use

$$S(q) = \frac{[f(|\mathbf{q}|)]^2}{N} \sum_{i,j} \langle \sigma_i \sigma_j \rangle (\mathbf{e}_i^\perp \cdot \mathbf{e}_j^\perp) e^{\mathbf{q}\cdot r_{ij}}$$

where the $<\sigma_i\sigma_j>$ are the correlations between Ising spins at sites $i$ and $j$, $\mathbf{e}_i^\perp$ is the component of the quantisation direction at site $i$ perpendicular to the scattering vector $\mathbf{q}$, $N$ is the number of spins and $f(|\mathbf{q}|)$ is the magnetic form factor for $Dy^{3+}$ extracted from [6].

We find that the $S(q)$ calculated from our model is compatible with the existing experimental results. It is important to remark that no effort was put before hand to enforce this compatibility: the $J$ values were optimised in relation to the double feature observed in the susceptibility measurements that we describe in the text.

The Supplementary Figure 3 shows a comparison between the experimental data from [7] at $T$ = 300 mK and our calculations for the same temperature in a lattice of $L = 6$ and over 200 averaged copies, each obtained after a zero field cooling:

Some noteworthy points of comparison are: i) The presence in both the experimental data and the simulation of hexagonal loops of diffuse scattering running along the Brillouin zone boundaries. ii) The presence of pinch points characteristic to spin-ice and iii) The presence of regions of intense scattering around (0,0,3) and (3/2, 3/2, 3/2). The narrow Bragg peaks, due to the lattice structure, are absent in our simulations.

## Supplementary References